# Ultrahigh-*Q* on-chip silicon-germanium microresonators

Ryan Schilling*, Chi Xiong, Swetha Kamlapurkar, Abram Falk, Nathan Marchack, Stephen Bedell, Richard Haight, Christopher Scerbo, Hanhee Paik, and Jason S. Orcutt

*IBM Quantum, T.J. Watson Research Center, 1101 Kitchawan Road, Yorktown Heights, NY 10598, USA*
*ryan.schilling@ibm.com*

**We demonstrate fully crystalline, single-mode ultrahigh quality factor integrated microresonators comprising epitaxially grown $Si_{0.86}Ge_{0.14}$ waveguide cores with silicon claddings. These waveguides support resonances with internal $Q > 10^8$ for both polarization modes, a nearly order-of-magnitude improvement over that seen in prior integrated Si photonics platforms. The maximum $Q$ is $1.71 \pm 0.06 \times 10^8$ for the transverse magnetic (TM) polarization mode, corresponding to a loss of $0.39 \pm 0.02$ dB/m. Together with silicon's strong Kerr nonlinearity and low losses in the optical, microwave and acoustic regimes, our results could lead to the $Si_{1-x}Ge_x$/Si architecture unlocking important new avenues for Kerr frequency combs, optomechanics, and quantum transduction**.

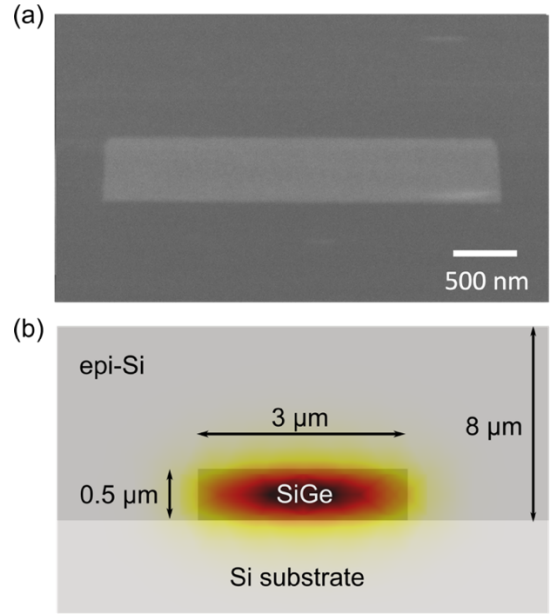

**Fig. 1.** (a) SEM of $Si_{0.86}Ge_{0.14}$ resonator core surrounded by epitaxial silicon. The corresponding dimensions of the waveguide and cladding layer with overlaid finite element model of TE mode electric field are illustrated in (b).

Whispering gallery mode resonators, which use continuously curved loops to guide electromagnetic waves by total internal reflection, have been extensively studied for more than half a century due to their ability to support very high quality ($Q$) factors. The highest $Q$ resonators have been fabricated with macroscopic machining techniques, such as manually polishing Si [1] and melting silica fibers into spheres [2], resulting in $Q > 10^9$ in both cases. An extensive effort is now underway to microfabricate resonators that can match or even exceed these $Q$ values. Indeed, in amorphous systems recent progress has significantly narrowed the gap between the quality of macroscopic and microfabrication processes. The best results for integrated silica resonators use wet etching to achieve $Q = 2.1 \times 10^8$ [3]. Recent progress in silicon nitride devices has utilized high temperature annealing to demonstrate $Q = 4.2 \times 10^8$ [4]. However, there remains a large gap in the performance of Si resonators. Despite much effort invested in silicon photonics, the best $Q$ achieved in silicon resonators has been $Q = 2.2 \times 10^7$ and requires a silica top cladding [5].

The many benefits of integrated optical resonators, including their capability of forming more complicated optical circuits and their compatibility with proximate electrodes, has led to their use in many applications [6–11]. Typically, these applications also demand high intrinsic $Q$ factors, such as for Kerr combs [12] and microwave-optical quantum transducers [13]. Silicon is particularly well suited to many of these applications due to its high refractive index, strong Kerr nonlinearity, broad infrared transparency, low acoustic loss, low microwave loss and compatibility with CMOS fabrication techniques. For that reason, there is broad interest in improving the performance of on-chip silicon resonators. By significantly improving the $Q$ factors of silicon-based integrated photonic resonators in this work, we are taking a major step in bridging the disparity in optical loss between microfabricated amorphous dielectric and crystalline silicon systems.

Our fully monocrystalline and epitaxial waveguide platform comprises a low mole-fraction silicon-germanium core that is fully clad by silicon. Optical losses due to pinned surfaces resulting in free carrier and/or defect state absorption are eliminated by extending the single-crystalline region throughout the region occupied by the optical mode. Furthermore, scattering losses are minimized by the low index contrast of our $Si_xGe_{1-x}$ waveguide relative to its Si cladding (an order of magnitude lower than that of Si relative to $SiO_2$).

Although $Si_xGe_{1-x}$ waveguides have attracted attention for linear and nonlinear applications [14,15], the best propagation loss in a $Si_xGe_{1-x}$ waveguide at telecom keyholing in narrow waveguide gaps, which arises due to crystallographic orientation-dependent growth rates [20]. Careful control over the growth conditions is critical in order to maintain strain in the metastable films (see S5). The chips are then either cleaved or diced from the wafer to expose the waveguide facets. All devices studied in this work employ a ring resonator with a radius of 4.8 mm, coupled to two (through/drop) waveguides with the same coupling gap.

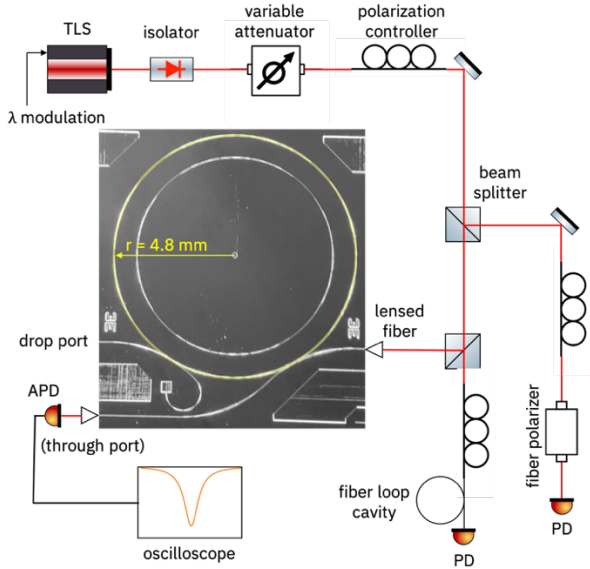

**Fig. 2.** Characterization setup used to measure transmission spectra around resonance at both through and drop ports. The input waveguide is seen at the right side lensed fiber and the through port is seen at the left side lensed fiber (drop port is above). The ring resonator is highlighted in yellow.

Characterization is carried out using the setup shown in Fig. 2. A lensed fiber on a precision stage couples light into the input waveguide, and a second lensed fiber collects the light either at the through or drop ports. Three wideband tunable diode lasers are used to collect the transmission port data (Figs. 3 and 4) over the range of 1300 nm to 1640 nm. For each measurement, the laser is tuned to the resonance wavelength of interest and a function generator is used to modulate the laser cavity length in order to perform a narrow continuous sweep across the resonance. Light is split off from the input fiber to a fiber loop cavity with a calibrated free-spectral range (FSR), which serves as a frequency reference for the narrow scan. Light is also split off to a pair of crossed polarizers that allows for tracking of the input polarization. An example of this type of scan is shown in Fig. 3(b), where the input polarization has been tuned to the TM mode. Here, the through and drop port data have been fitted wavelengths was reported to be 65 dB/m in $Si_{0.9}Ge_{0.1}$ waveguides [16]. In contrast, our optimization of waveguide loss focuses on maintaining a fully strained $Si_xGe_{1-x}$ layer, which is achieved through experimental determination of the maximum metastable thickness, as well as minimizing thermal shock during and thermal budget during the epitaxial silicon cladding growth. In addition, we focus on minimizing processing contamination that can lead to interface defects. We are thus able to achieve a loss rate of $0.39 \pm 0.02$ dB/m. With a fully strained $Si_xGe_{1-x}$ waveguide core and epitaxial silicon cladding, the core/cladding index contrast can be optimized, and defect-state absorption minimized.

Our waveguides are defined in epitaxial $Si_{0.86}Ge_{0.14}$ layers grown on high resistivity Cz-Si wafers. An important performance parameter is the $Si_{0.86}Ge_{0.14}$ thickness. If it is too thin, radiation loss will affect the performance. On the other hand, when $Si_{0.86}Ge_{0.14}$ is grown beyond the Matthews-Blakeslee critical thickness [17,18], plastic relaxation emerges that can degrade the strain and create defects. Thus, our target thickness for $Si_{0.86}Ge_{0.14}$ is near the expected metastable critical thickness observed in prior work [19]. The grown layers are determined to be fully strained within the precision of X-ray diffraction (XRD) analysis and then more precisely evaluated with defect-selective etching (see S5). Waveguides are then defined in the $Si_{0.86}Ge_{0.14}$ by optical lithography followed by inductively coupled plasma (ICP) etching. The surface is then cleaned, and an epitaxial silicon cladding layer is grown on top, with a controlled ramp rate and minimum growth temperature to reduce relaxation of the $Si_{0.86}Ge_{0.14}$. A dichlorosilane (DCS) precursor is chosen for the epitaxial silicon top cladding to prevent in order to extract the internal $Q$ of the ring resonator itself. In this case an internal $Q = 1.71 \pm 0.06 \times 10^8$ is measured for the TM mode at 1450 nm (see S3 for $Q$ extraction). This characterization was carried out for modes across a broad spectral range, for TM and TE polarizations (Fig. 4(a) and (b)). For both, $Q$ was maximized near 1450 nm, with a trend of decreasing $Q$ at both shorter and longer wavelengths. The increasing loss at shorter wavelengths is attributed to the expected effect of band edge absorption. At longer wavelengths, the dominant loss mechanism for the TE mode is radiation deriving from to the bending of the waveguide – an effect known as radiation loss. The TM mode, which exhibits a higher refractive index due to the birefringence of strained $Si_xGe_{1-x}$, is not radiation loss-limited.

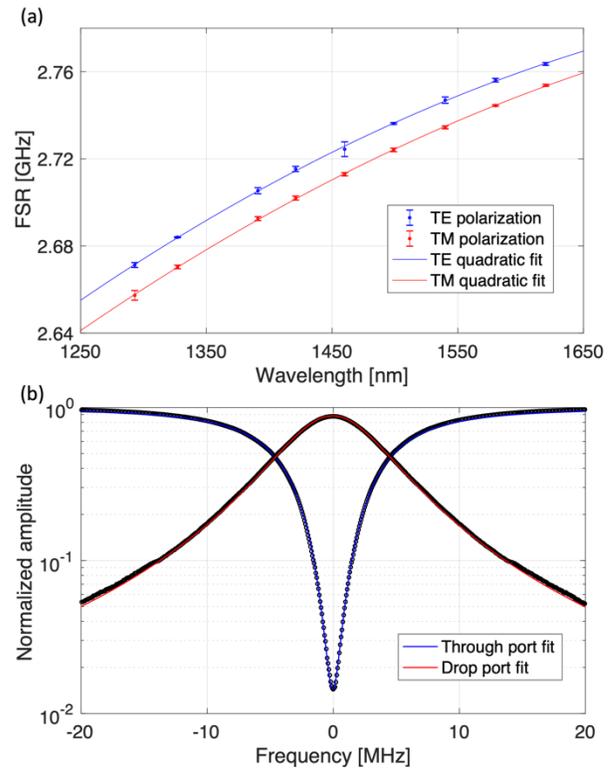

**Fig. 3.** (a) Extracted free-spectral ranges for both polarizations as a function of wavelength (one standard deviation error bars). Solid lines show a quadratic fit to the data. (b) Zoom-in and fitting to the through and drop port data to determine intrinsic loss rate of the TM mode at 1450 nm.

The radiation loss model plotted in Fig. 4(b) utilizes a single fitting parameter, which is the $Si_{0.86}Ge_{0.14}$ refractive index, $n_{SiGe}$ (see S4). The geometry of the modelled waveguide is fixed by the measured parameters of the physical waveguide. The fitted result for the in-plane (TE) index, infers $n_{SiGe,TE} = 1.013 n_{Si}$. The close agreement between data and radiation loss-limited scaling in the L-band strongly suggests this is the dominant loss mechanism.

The birefringence between TE and TM polarizations in our system is inferred from the differing free-spectral ranges (FSR) shown in figure 3(a), combined with numerical mode simulation (see S4). The TM index is determined to be $n_{SiGe,TM} = 1.023 n_{Si}$. These results agree fairly well with the result $n_{SiGe} = 1.0115 n_{Si}$ that was obtained through ellipsometry (S1), using an isotropic model for the refractive index. We determine that $n_{SiGe} = 1.015 n_{Si}$ is sufficient for the radiation loss-limit to exceed the $Q$s measured in Fig. 4(a).

A previous study of the polarization dependent refractive index for pseudomorphic $Si_xGe_{1-x}$ waveguides grown on Si also observed an anisotropy between the out-of-plane and in-plane indices [21]. This work, based on experimentally tracing out the mode profiles, inferred an index difference of $\delta n_{TM} - \delta n_{TE} = (0.21 \pm 0.07)x$, for strained $Si_{1-x}Ge_x$ films.

For $Si_{0.86}Ge_{0.14}$ studied in this work, the index difference is experimentally determined from quadratic fits to the polarization dependent FSRs. Taking the difference of the quadratic fits, allows us to establish the birefringence in terms of the ratio of effective group indices, $n_{g\,eff,TM}/n_{g\,eff,TE}$ (see Fig. S7). Using the value for $n_{SiGe,TE}$ determined from the radiation loss model and a finite difference eigenmode (FDE) solver to translate between real and effective index, $n_{SiGe,TM}$ is determined (see Fig. S8). The result gives a birefringent index difference of $\delta n_{TM} - \delta n_{TE} = 0.24x$, in close agreement with prior literature.

As can be seen in Fig 4(b), the agreement between experiment and radiation-limited $Q$ does not hold at wavelengths below 1500 nm for the TE mode. In fact, at 1450 nm, the limit to $Q$ imposed by curvature-derived radiation is nearly an order of magnitude higher than experiment. As discussed, the TM data is not radiation-limited at any measured wavelength. Therefore, neither loss mechanism -- neither band edge absorption nor radiation loss – is setting the $Q$ limit in the E- and S-bands. Although the mechanism is not identified here, a likely candidate is scattering at the etched $Si_{0.86}Ge_{0.14}$ / Si interface. At longer wavelengths the $Q$ may be degraded as a result of strong coupling to the bus waveguides leading to hybridization.

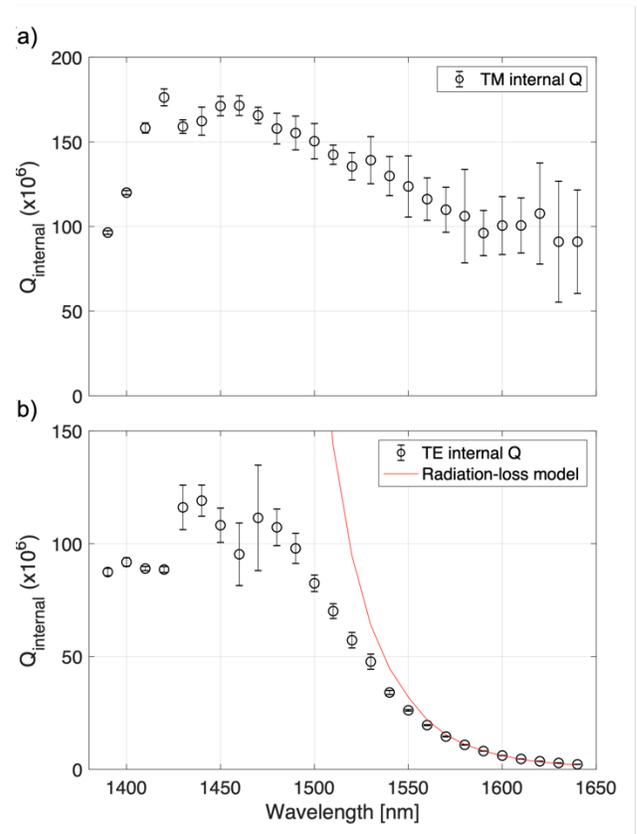

**Fig. 4.** Provides a comparison of experimentally measured optical $Q$ as a function of wavelength, for (a) TM and (b) TE polarizations in a single mode microresonator (open circles, with one standard deviation error bars). Overlaid with a solid red line for TE, is a radiation loss model utilizing the $Si_{0.86}Ge_{0.14}$ refractive index as a single fitting parameter.

One of the main features of the radiation loss regime is the extreme sensitivity of $Q$ to material and design parameters. For instance, in our regime, the calculated radiation-limited $Q$ improves by three orders of magnitude by doubling $R$ from 4.8 mm to 9.6 mm. Alternatively, increasing the $Si_{0.86}Ge_{0.14}$ thickness from 500 nm to 600 nm would result in an order of magnitude lower radiative loss in the C-band.

We have demonstrated an ultrahigh quality factor on-chip silicon-germanium microresonator system with unprecedented low loss rates. It uses a novel fully crystalline design based on the low refractive index contrast between Si and $Si_{0.86}Ge_{0.14}$ to minimize interface scattering losses. Relative to the prior state-of-the-art in silicon ring resonators [5], our results improve waveguide loss by nearly an order of magnitude. Moreover, there is clear space to improve substantially on these current results by optimizing the Si$_x$Ge$_{1-x}$ growth process and exploring other Ge mole fractions. Currently, our $Q$ values are nearly on par with the best on-chip, waveguide coupled microresonators, which are formed by wet etching of silica [3]. Moreover, our system has the added advantage of being single mode, allowing precise and reproducible control over dimensions by dry etching, being solid and thus immune to mechanical vibrations, and being hermitically sealed from the environment by way of an encapsulating silicon layer.

The Si$_x$Ge$_{1-x}$ system is also particularly well suited to build microwave-optical transducers for quantum state transfer [22] In addition to the low optical losses demonstrated here, this epitaxial system has very low microwave losses, as has been recently demonstrated by the fabrication of transmon qubits with > 100 us coherence times on similar substrates [23]. Moreover, it is also expected that the low acoustic loss of silicon is expected to make this a favorable direction for optomechanics-based transducers [24,25]. Beyond transducers, the favorable properties of a platform with extraordinarily low losses in the optical, microwave, and acoustic regimes should enable figure of merit improvements for a wide variety of applications, including compact spectrometers, precision mass sensors and optically generated microwave sources.

## Acknowledgments

We acknowledge funding under the Army Research Office / Laboratory for Physical Sciences Cross-Quantum Technology Systems Contract Number W911NF-18-1-0022. The authors wish to acknowledge support in device fabrication from the IBM Microelectronics Research Laboratory and IBM Central Scientific Services.

# Supplement

## 1. Device Fabrication

Epitaxial Si$_{0.86}$Ge$_{0.14}$ layers with a thickness of 500 nm with a 20 nm capping silicon layer were grown on (100) high-resistivity 200 mm Si substrates, by reduced-pressure chemical vapor deposition chamber (RP-CVD) with silane and germane precursors. Next, the lithographic mask stack was established using the following steps: low-pressure chemical vapor deposition (LPCVD) of 100 nm silicon dioxide (SiO$_2$) at 400°C, followed by spin coating of 200 nm organic planarizing layer (OPL), 35 nm silicon-containing antireflective coating (SiARC) and 70 nm of photoresist (PR). The waveguide pattern was defined in the PR layer by exposure to an ArF 193 nm wavelength excimer laser followed by development in an n-butyl acetate solution. Dry etching of the SiARC, OPL and SiO$_2$ mask layers was conducted in an inductively coupled plasma (ICP) chamber, followed by a strip process to remove residual OPL in a downstream microwave (MW) plasma chamber, and ending with Si$_{0.86}$Ge$_{0.14}$ etching in the ICP chamber. Wafers were transferred between the ICP and MW reactors without breaking vacuum. The SiO$_2$ mask was removed from the etched Si$_{0.86}$Ge$_{0.14}$ waveguides using a 10% dilute hydrogen fluoride (DHF) solution, followed by piranha solution (5:1 H$_2$SO$_4$ : H$_2$O$_2$) to remove any organic remnants. Following this, 8 μm of epitaxial silicon was grown over the waveguides at 800°C using dichlorosilane as precursor with a controlled temperature ramp from 450°C, at a rate of 1°C/s.

## 2. Ellipsometry of SiGe

In order to experimentally measure $n_{SiGe}$, we used a J.A. Woolam variable-angle spectroscopic ellipsometer (Fig. S2). Specifically, we measured films comprising 500 nm (nominally) of epitaxially grown Si$_{0.86}$Ge$_{0.14}$ capped by 8 μm of Si. The resulting measured amplitude ratio (tan Ψ) and phase difference (Δ(°)) show doubly periodic oscillations, which derives from reflections at the Si$_{0.86}$Ge$_{0.14}$ – Si interfaces and the top surface of the chip.

In addition to the fitting parameters $A$ and $B$, we also allowed the film thicknesses to be fitting parameters. We used the Woolam CompleteEASE software to model the results. With the Woolam-provided Si index of refraction being $n_{Si-JAW}$, we modelled the stack as having the following layers:

Layer 1, Silicon: $n = n_{Si-JAW}$

Layer 2, Si$_{0.86}$Ge$_{0.14}$: $n = n_{Si-JAW} + A$, where $A$ is a λ–independent fitting parameter

Layer 3, epitaxial Si: $n = n_{Si-JAW} + B$, where $B$ is a λ-independent fitting parameter

Layer 4, native SiO$_2$: $h = n_{SiO2-JAW}$

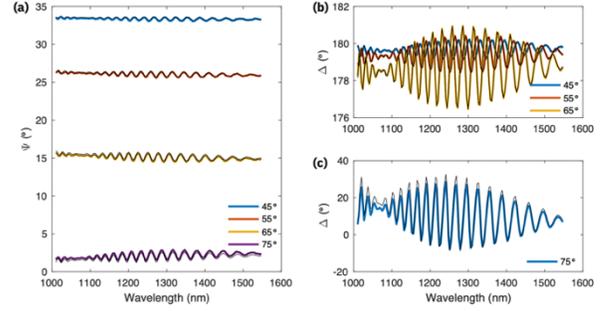

**Fig. S1.** Ellipsometry data for Si$_{0.86}$Ge$_{0.14}$, showing **(a)** the measured amplitude ratio **(b)** and phase difference with models overlaid.

We find that for our Si$_{0.86}$Ge$_{0.14}$ films at l = 1550 nm, $n_{SiGe} = 1.0115 n_{Si}$. This is in fairly good agreement with the $n_{SiGe,TE} = 1.013 n_{Si}$ fitted through our radiation loss calculations.

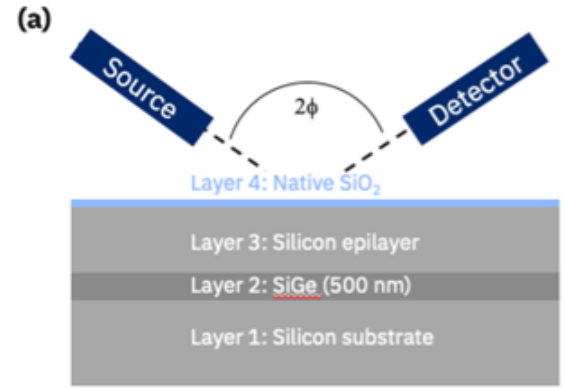

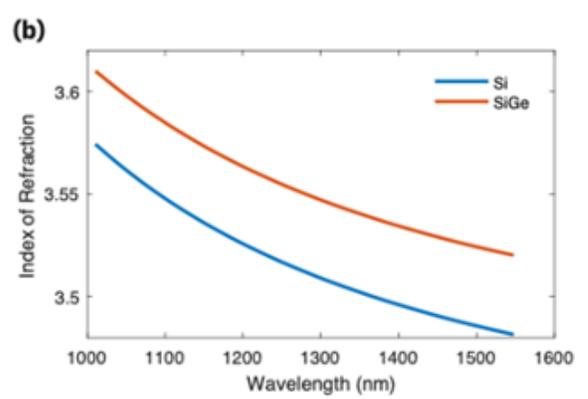

**Fig. S2.** **(a)** Ellipsometry setup and film stack that was measured and modelled. **(b)** The fitted result $n_{Si0.86Ge0.14}$, with $n_{Si-JAW}$ also plotted as a comparison.

## 3. Q Extraction

This section describes the analysis used to determine the intrinsic Q of the ring resonators studied in this work. The rings are symmetrically coupled to two waveguides, in an add-drop configuration, as shown in Fig. S3.

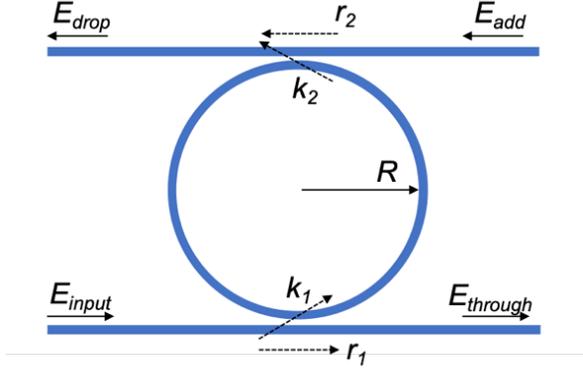

**Fig. S3.** Add-drop configuration for coupling to a ring resonator.

One waveguide serves to couple light from the CW laser source into the ring, via the input port. The degree to which input light is split, between the through port and ring, at the coupling region is defined by the cross-coupling parameter, $k_1$, and self-coupling parameter, $r_1$, respectively. We assume that no loss occurs at the coupling region, such that the sum of the power splitting ratios is unity, $r_1^2 + k_1^2 = 1$. Likewise, for the add-drop waveguide $r_2^2 + k_2^2 = 1$. This configuration is analogous to a Fabry-Perot interferometer, where the electric field at the through port can be defined in terms of the input field as follows:

$$E_{through} = \left[ r_1 + \frac{r_2 k_1 k_2 a e^{i\phi(r_1^2 + k_1^2)}}{1 - r_1 r_2 a e^{i\phi}} \right] E_{input}$$

where $a$ represents the round-trip loss in the ring and $\phi$ is the round-trip phase shift. Similarly, the field intensity at the drop port is given by:

$$E_{drop} = \left[ \frac{k_1 k_2 \sqrt{a} e^{i\phi/2}}{1 - r_1 r_2 a e^{i\phi}} \right] E_{input}$$

In order to extract the round-trip loss and cross-coupling parameters experimentally, we must re-write these equations in terms of the light intensity measured at the through and drop ports. We normalize to the input field and because of the symmetric coupling arrangement can simplify the analysis by assuming that $r_1 = r_2 = r$. This leads to the following expressions:

$$T_{through} = \left| \frac{E_{through}}{E_{input}} \right|^2 = \frac{r_2^2 a^2 - 2r_1 r_2 a \cos\phi + r_1^2}{1 - 2r_1 r_2 a \cos\phi + (r_1 r_2 a)^2}$$

$$\overline{r_1 = r_2 = r} \frac{r^2(a^2 - 2a\cos\phi + 1)}{1 - 2r^2 a \cos\phi + r^4 a^2}$$

$$T_{drop} = \left| \frac{E_{drop}}{E_{input}} \right|^2 = \frac{(1 - r_1^2)(1 - r_2^2)a}{1 - 2r_1 r_2 a \cos\phi + (r_1 r_2 a)^2}$$

$$\overline{r_1 = r_2 = r} \frac{(1 - r^2)^2 a}{1 - 2r^2 a \cos\phi + r^4 a^2}$$

The intrinsic quality factor can then be defined in terms of the ring resonator length, $L = 2\pi R$, group refractive index, $n_g$, and the resonance wavelength, $\lambda_{res}$:

$$Q_{intrinsic} = \frac{\pi n_g L \sqrt{a}}{\lambda_{res}(1 - a)}$$

The measurement proceeds by scanning a tunable diode laser over a small frequency range, around a resonance of interest and recording the photodiode signal from one of the output ports, using an oscilloscope. Part of the light is split off and passed through a fiber loop cavity. This process is done for both the through and drop ports. The resulting data set for a single wavelength is shown in Fig. S4. Calibration of the fiber loop cavity was performed using an Agilent 81600B tunable laser source, with a build-in wavelength meter. The free-spectral range was found to be 97.6 MHz at 1450 nm.

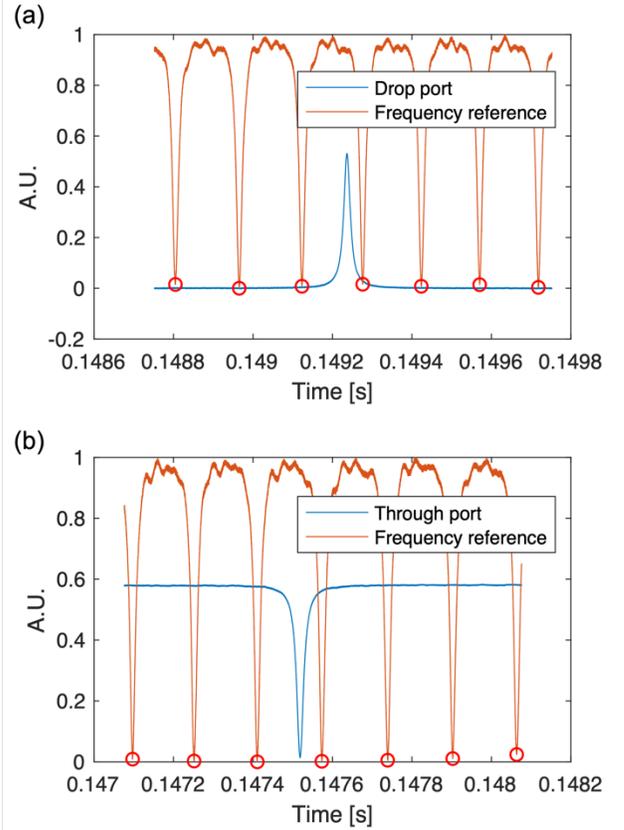

**Fig. S4.** Unprocessed oscilloscope traces, showing the drop/through port (blue) and frequency reference (red) measurements.

For each measurement, the photodetector background level is also measured and subtracted from the trace, which is then normalized. The resonance peaks of the frequency reference are located and used to transform the data into frequency domain. It is then converted into phase domain using the known ring radius and effective refractive index. The equations above for $T_{through}$ and $T_{drop}$ are fitted using a trust-region-reflective nonlinear least-squares solver, to simultaneously minimize the residuals of both functions. An example of the fitted data is shown in Fig. 3. This measurement is repeated ten times at each wavelength to establish a mean value and error bounds.

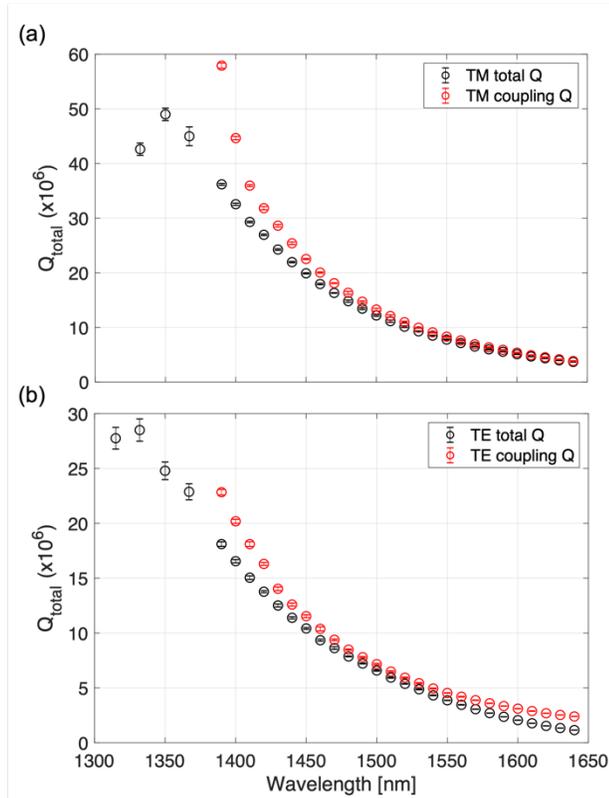

**Fig. S5.** Total (black) and coupling (red) $Q$-factors for the (a) TM and (b) TE polarizations of the exemplar sample studied in this work (open circles, with one standard deviation error bars).

It is instructive to consider the relationship between $Q_{total}$ and $Q_{coupling}$, as plotted in Fig. S5, where $Q_{coupling}$ is the $Q$ limit imposed by losses to the waveguides. For both polarizations it can be observed that where the two $Q$ values tend towards convergence, error bounds on the fitted $Q_{intrinsic}$ (Fig. 4) become largest. This is due to near complete extinction on the through port. In this regime the internal $Q$ estimate is highly sensitive to detector and polarization noise, and this results in a larger degree of uncertainty. At shorter wavelengths stronger field confinement in the core results in external loss to the waveguides becoming less significant. At the same time, $Q_{intrinsic}$ is also reduced at these wavelengths, as $Si_{0.86}Ge_{0.14}$ band edge absorption increases. At the shortest wavelengths measured we expect absorptive losses to become dominant. At the longest wavelengths, for the TE mode, coupling loss to the waveguides increases but radiative loss from the ring increases much more rapidly. However, in the range around 1450 nm it appears that there is an optimal trade-off between radiation and absorption loss mechanisms. In this regime, coupling losses dominate, leading to a large discrepancy between the intrinsic and total $Q$ factors. This is the range where the TE mode's estimated $Q_{intrinsic}$ has the highest uncertainty. However, that is not the case for the TM mode, which has weaker coupling to the waveguides and thus a higher $Q_{coupling}$. For the TM polarization, it is at wavelengths beyond 1500 nm that $Q_{coupling}$ becomes commensurate with $Q_{total}$ and the fitting results become most uncertain.

### 4. SiGe Refractive Index Anisotropy

A refractive index contrast between $Si_{0.86}Ge_{0.14}$ and Si of 1.31% for the TE mode is determined from the radiation loss fit on the L-band data in figure 4(b). In order to model the radiation loss, we use a conformal mapping procedure [26], whereby the stepwise flat index profile of a curved waveguide is mapped to a graded index of a straight waveguide. Calculating the loss rate in the latter amounts to a one-dimensional tunneling problem and can then be mapped back to the former. The result is an almost entirely analytic solution for the radiation-limited $Q$. The one component of this calculation that needs to be calculated numerically is the effective index of the optical modes. To do this calculation, we use the Lumerical MODE 2D finite difference eigenmode (FDE) solver.

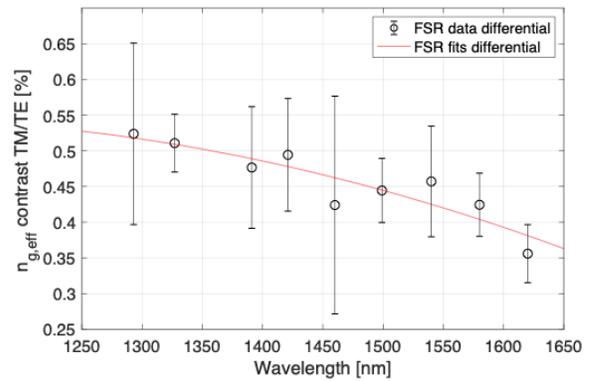

**Fig. S7.** Complementary plot to main figure 3(a), showing the polarization-dependent difference in effective group index, inferred from the ratio of FSRs, as a function of wavelength. The red line plots this relation, taken from the ratio of quadratic fits to the FSR data in figure 3(a).

The same radiation loss model predicts that a refractive index anisotropy of 0.2% is sufficient to put radiation loss-limited $Q$s in excess of the L-band data for the TM polarization in figure 4(a). The experimentally determined birefringent anisotropy of 0.99% confirms that the TM polarization mode is not radiation loss-limited in the range studied in this paper.

In order to determine the birefringence of the Si$_{0.86}$Ge$_{0.14}$, the ratio of effective group indices, $n_{g\,eff,TM}/\delta n_{g\,eff,TE}$, was determined experimentally (Fig. S7). Using an FDE solver to determine the relation between $n_{g\,eff}$ and $n_{SiGe}/n_{Si}$ allowed for the material birefringence to be extracted. The birefringent index difference was found to be $\delta n_{TM} - \delta n_{TE} = 0.24x$, with $n_{SiGe,TE} = 1.013 n_{Si}$ determined from the radiation loss model fit to the L-band TE $Q_{intrinsic}$ data. Using literature values for Si dispersion, the dispersion plot in Fig. S8 was produced.

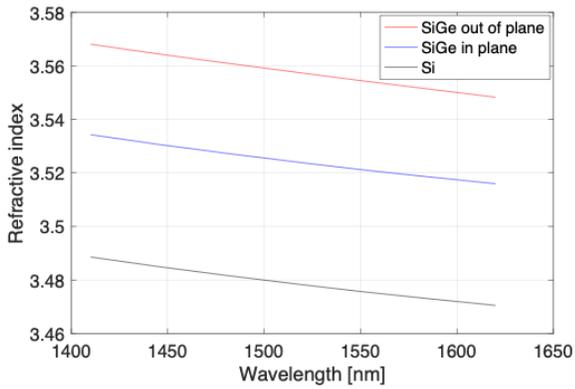

**Fig. S8.** Birefringent dispersion of Si$_{0.86}$Ge$_{0.14}$ determined from radiation loss fit to the TE $Q_{intrinsic}$ data and polarization-dependent FSR data.

## 5. Epitaxy of SiGe and Si

The Si$_{0.86}$Ge$_{0.14}$ films utilized in this paper were grown to be as thick as possible, while remaining pseudomorphic. Specifically, we aimed to increase the layer thickness until the point where plastic relaxation begins to occur but does not yet lead to a significant reduction in the layer strain. Beyond this point, relaxation of the Si$_{0.86}$Ge$_{0.14}$ leads to reduction in the refractive index and thus reduced confinement of the optical mode.

Before growth commences it is necessary to prepare the Si surface by removing the native oxide layer (typically several nm in thickness). This is accomplished by an in-situ H$_2$ bake at 1050 °C. This leads to the reduction of the oxide layer, resulting in a clean Si surface ready for epitaxy. The chamber temperature is then reduced to approximately 650 °C, for Si$_{0.86}$Ge$_{0.14}$ growth.

During growth, when the Si$_{0.86}$Ge$_{0.14}$ becomes sufficiently thick, the stress induced by the lattice mismatch between Si and Si$_{0.86}$Ge$_{0.14}$ can lead to dislocation propagation. These dislocations move through the crystal along the <111> planes, which relieves some of the layer strain. On (100) wafers, this appears as surface cross-hatching where the density of the lines and magnitude of the topography increases as the layer thickness increases. The nucleation of dislocations tends to originate from defects at the Si/Si$_{0.86}$Ge$_{0.14}$ growth interface, predominantly at the edge of the wafer. Therefore, the density of cross-hatching will tend to be largest at the edge and to decrease towards the center. They can be easily identified by highlighting with a defect selective etch chemistry, such as Secco ((K$_2$Cr$_2$O$_7$-H$_2$O)/HF). This approach is shown in Fig. S9, where a region near the edge of the wafer and a region at the center are compared. Near the edge a high density of cross-hatching lines are visible but at the center only a single line is visible. Also visible in the images are the dislocation cores that thread to the surface of the crystal and appear as pits after defect etching. These can also be identified using atomic force microscopy or laser scattering techniques.

Although the cross-hatching line density may be high near the wafer edge, this is not a concern for our process as the waveguide etch effectively isolates these defective regions from more pristine areas closer to the center. So, in optimizing our growth conditions we focus on the center region of the wafer in order to maximize the Si$_{0.86}$Ge$_{0.14}$ thickness.

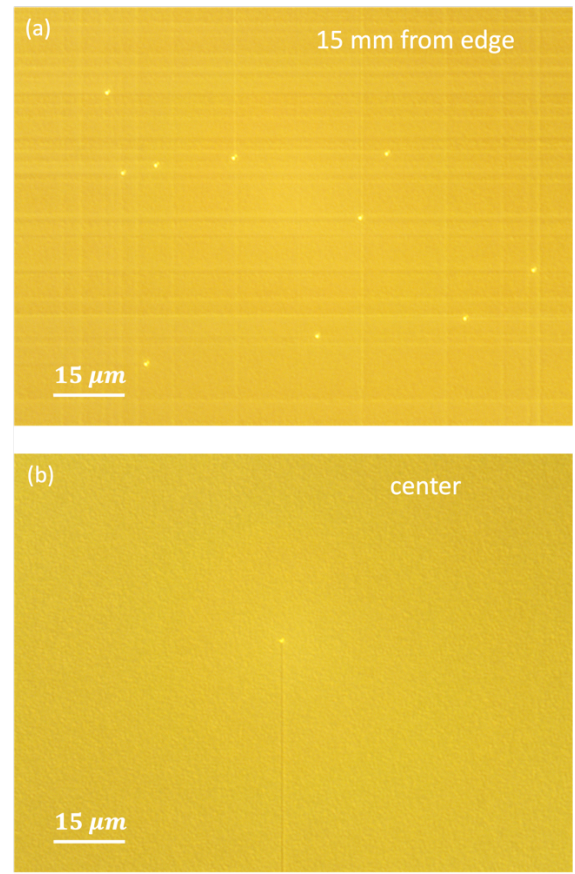

**Fig. S9.** Surface cross-hatching, and individual dislocation cores, seen on a Si$_{0.86}$Ge$_{0.14}$ layer grown on a 200 mm wafer, (a) near the edge and (b) at the center, after defect etching.